\documentclass[prb,10pt,twocolumn,floatfix]{revtex4-1} 
\usepackage{amsmath}  % needed for \tfrac, \bmatrix, etc.
\usepackage{amsfonts} % needed for bold Greek, Fraktur, and blackboard bold
\usepackage{graphicx} % needed for figures
\usepackage{mhequ}
\usepackage{lipsum}

\def\condblank{\futurelet\thenext\thecomma}
\def\ie{{\it i.e.,}\condblank}

\def\point#1#2#3{\rlap{\kern #1 truecm
    \raise #2 truecm \hbox{#3}}}
\def\thecomma{\ifx,\thenewxt \else\ifx;\thenext \else\ifx.\thenext
        \else\ifx!\thenext \else\ifx:\thenext\else\ifx)\thenext \else \
        \fi\fi\fi\fi\fi\fi}
\def\fref#1{Fig.~\ref{#1}}
\def\eref#1{Eq.~(\ref{#1})}

\let\phi=\varphi
\def\p{{\bf p}}
\def\x{{\bf x}}
\def\t{{\bf t}}
\def\d{{\rm d}}
\usepackage{url}
\makeatletter
\g@addto@macro{\UrlBreaks}{\UrlOrds}
\makeatother

\begin{document}
\title{Broken Pencils and Moving Rulers:\\
  After an unpublished book by Mitchell Feigenbaum}

  \author{Jean-Pierre Eckmann}
\affiliation{D\'epartement de Physique Th\'eorique and Section de
  Math\'ematiques, University of Geneva, Geneva, Switzerland}\date{\today}
\begin{abstract}
Mitchell Feigenbaum discovered an intriguing property of viewing
images through cylindrical mirrors or looking into water. Because the
eye is a lens with an opening of about 5mm, many different rays of
reflected images reach the eye, and need to be interpreted by the
visual system. This has the surprising effect that what one perceives
depends on the orientation of the head, whether it is tilted or not. I
explain and illustrate this phenomenon on the example of a human eye
looking at a ruler immersed in water.

\end{abstract}
\maketitle
\section{Introduction: Anamorphic Images and Caustics}
Mitchell Jay Feigenbaum (Dec 19, 1944-June 30, 2019) was a well-known
mathematical physicist, whose work on period doubling\citep{feigenbaum1978,feigenbaum1979} is
known to many as one of the founding papers of the theory of
chaos.
Towards the end of his life, he worked intensely on a book whose
working title was ``Reflections on a Tube.''
\begin{figure}[h!]
\centering
\includegraphics[width=3in]{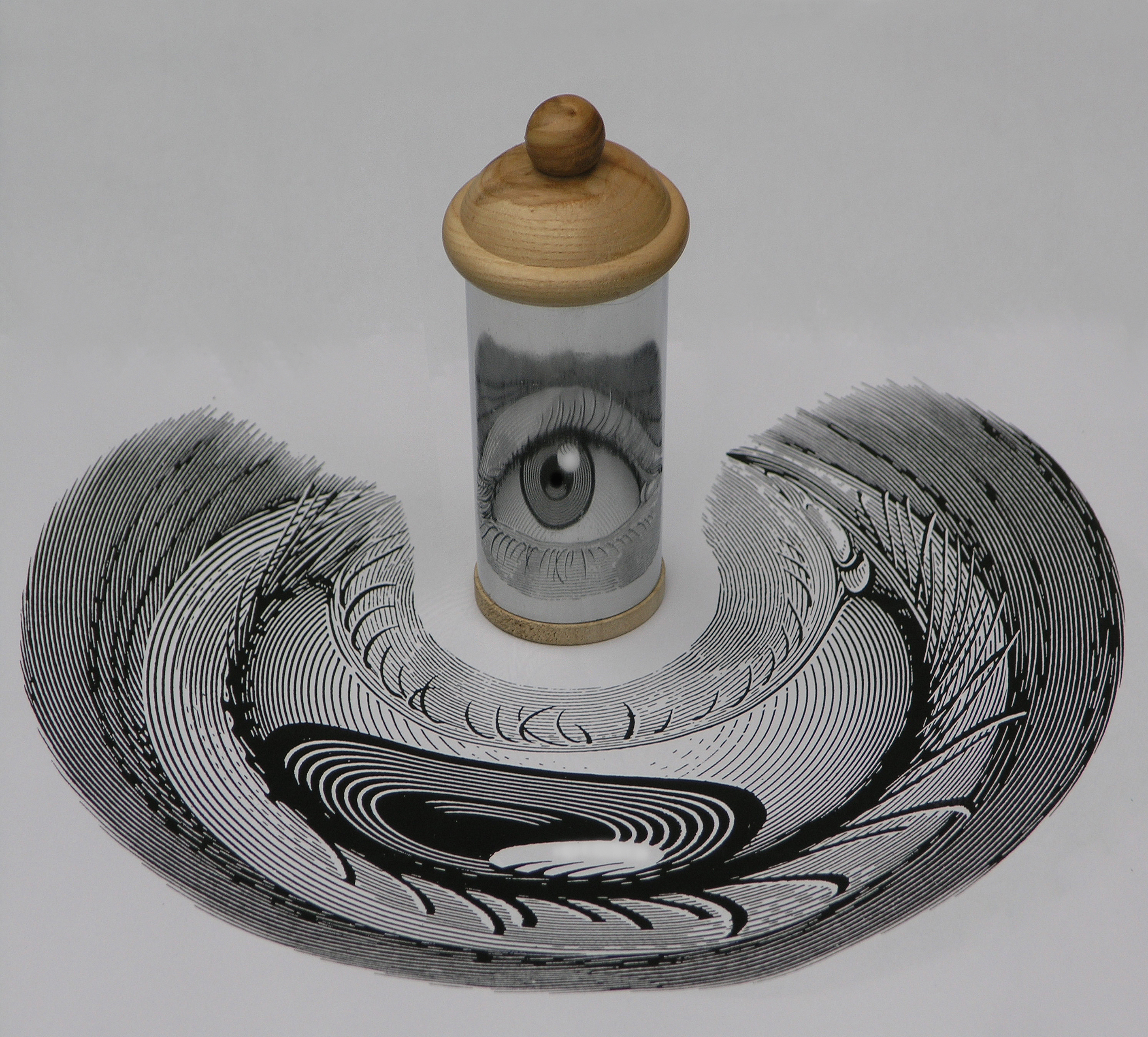}
\caption{(Color online) A modern anamorphic image, by Istv\'an
  Orosz. A very deformed eye is drawn in the plane, and shows as a
  undeformed eye in the cylinder.
  If you have such a
  cylinder, you will find out that the image is not on the surface of
  the cylinder, but on a surface with elliptical cross section inside
  the cylinder. This is called ``viewable surface'' by Feigenbaum. It
  is difficult to calculate this surface.}
\label{fig:anacyl}
\end{figure}

This book starts with a
study of anamorphic images, \ie what you see in a tube placed on a
table. The study of cylindrical mirrors was started in the
$17^{\textrm {th}}$ century, as soon
as people were able to make mirrors.\cite{niceron1638} I show a modern example of this in \fref{fig:anacyl}. Feigenbaum's study
starts with asking what one really sees when one views the image
reflected on the tube. Is it inside the tube, on the tube, or even
seen behind the tube? 

As this paper cannot do justice to all the ramifications these questions
generate, I will instead concentrate on a beautiful application
of the underlying principles and consider an experiment (also suggested in Feigenbaum's
book) which can easily be done with very limited equipment.

But let me start at the beginning. To understand what is at stake, let
me first have a look at the light reflected by a shiny cylinder. Each
point of the image in the plane below the cylinder emits light rays
in all directions, some of which hit the cylinder and get reflected
toward the eye of the observer.
How does one detect what is seen by the
eye?

A natural, but simplistic, idea is to draw a line from the eye to the
cylinder, and then, down to the paper, obeying the laws of reflection
(incoming angle to the tangent plane $=$ outgoing angle from that
plane). This method is commonly called ``ray-tracing''.\citep{rays}

What this idea overlooks is that the eye is not a pinhole camera and that
one should therefore consider \emph{all} the rays 
emanating from the source which reach the eye.

Many
\emph{different} rays will enter the eye through its opening (which is
about 5mm in the young adult). So what does the eye do with all these
rays coming   from just one point source? The eye measures
\emph{intensity}, and this intensity is maximal at the
\emph{caustic},\cite{wiki:caustic} which is the point where most rays
accumulate. In each direction, there is one such point, and these
points, together, form the viewable surface. (We will see later that
there are actually \emph{two} such points, giving rise to two vieable
surfaces.) 

\begin{figure}[h!]
\centering
\includegraphics[width=\columnwidth]{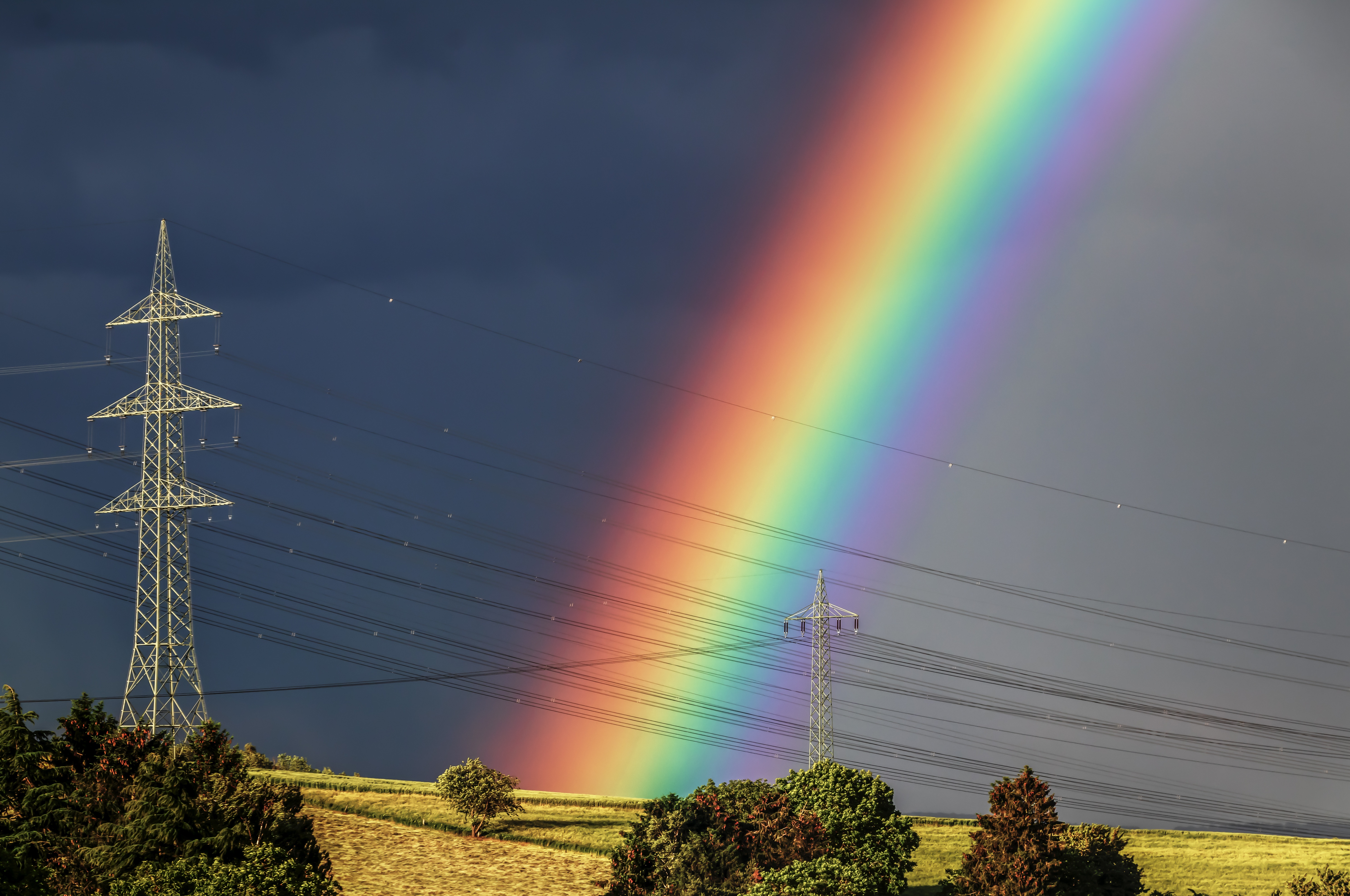}
\caption{(Color online) A photograph of a rainbow under optimum  conditions. Note that
  the sky is darker outside the rainbow than inside. Also note that the
  maximal intensity of each color is at the outside of the
  region of same color.
  Actually, looking from the center, the rainbow is a
  cross-section of the caustics.}
\label{fig:rainbow}
\end{figure}

The effects of caustics are well-known in rainbows, where you see maximal
intensity at the outer edge of each color\citep{minnaert,nussenzweig1977}
and a lighter background inside the rainbow as shown in \fref{fig:rainbow}.

\begin{figure}[h!]
  \begin{centering}
\includegraphics[width=\columnwidth]{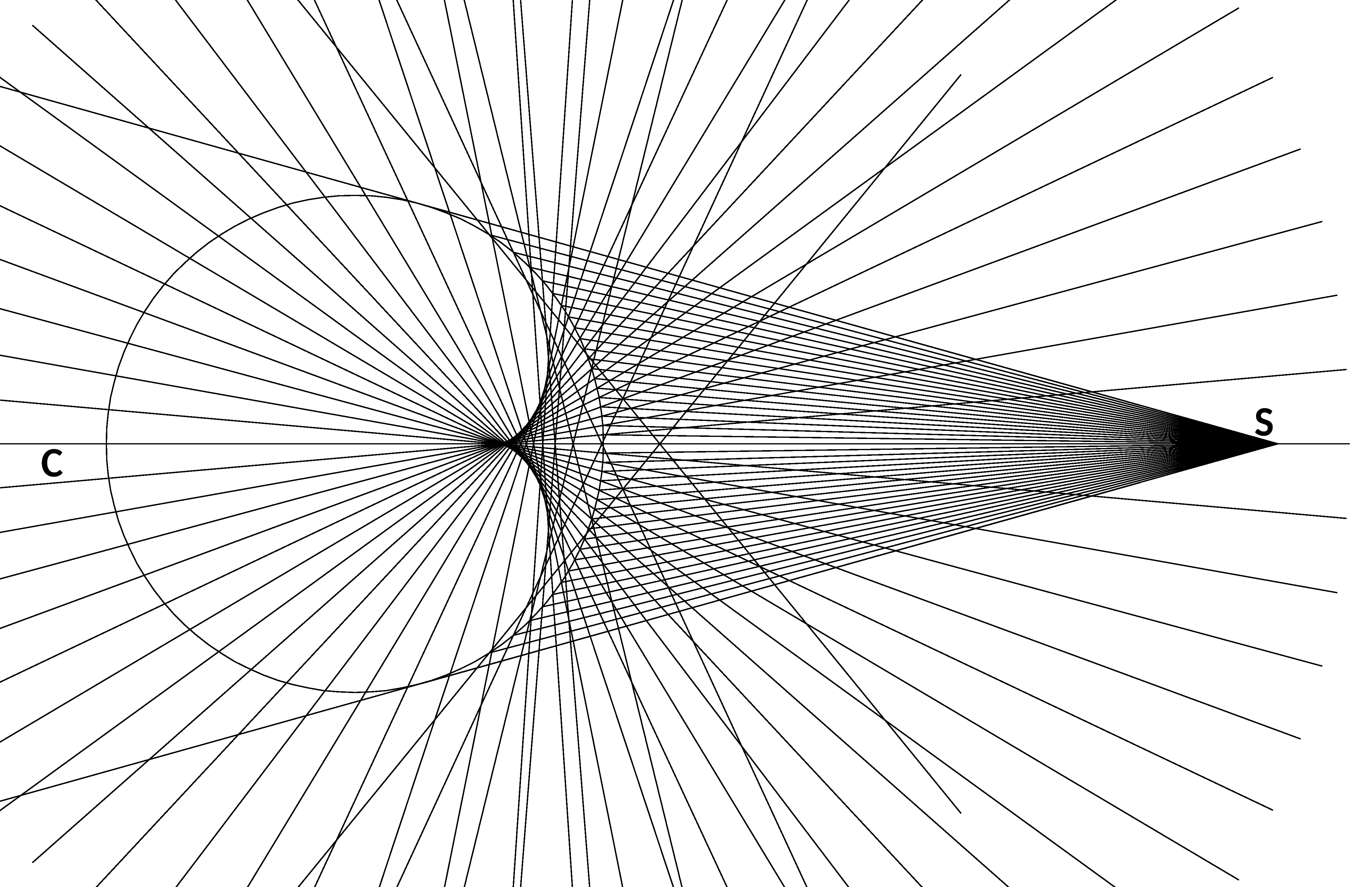}
  \end{centering}
  \caption{The caustics formed by light emanating from the source $S$
    and reflected on the circle $C$. The circle $C$ is the
    cross-section of the tube shown in \fref{fig:anacyl}. Note the
    cardioid-shaped caustic in the interior of the circle.}
  \label{fig:causbottom}
\end{figure}

To define the notion of caustic more precisely, look at \fref{fig:causbottom}.
Light is emitted from the source point $S$
and is reflected at the circle $C$. Each ray is reflected to an
outside ray which, in the drawing, is also continued \emph{inside}
the circle. The collection of rays forms a darker (cardioid)
curve. This curve is called the \emph{caustic}. The rays involved are
tangent to the caustic, and this defines it. The viewer perceives the
rays as coming from a bright source, the caustic, inside the circle. A
simpler but similar phenomenon can be
observed in any coffee cup, when light strikes the inside of the cup.\cite{wiki:cardioid}

\section{Looking at an immersed ruler}
The aim of this paper is to explain how the eye perceives caustics
when looking at a ruler which is immersed into water. This
problem is mathematically simpler than the one of the cylindrical
anamorphs. Furthermore, it is easy to make the experiment in a
classroom with minimal material.

\begin{figure}[h!]
\centering
\includegraphics[width=0.485\columnwidth]{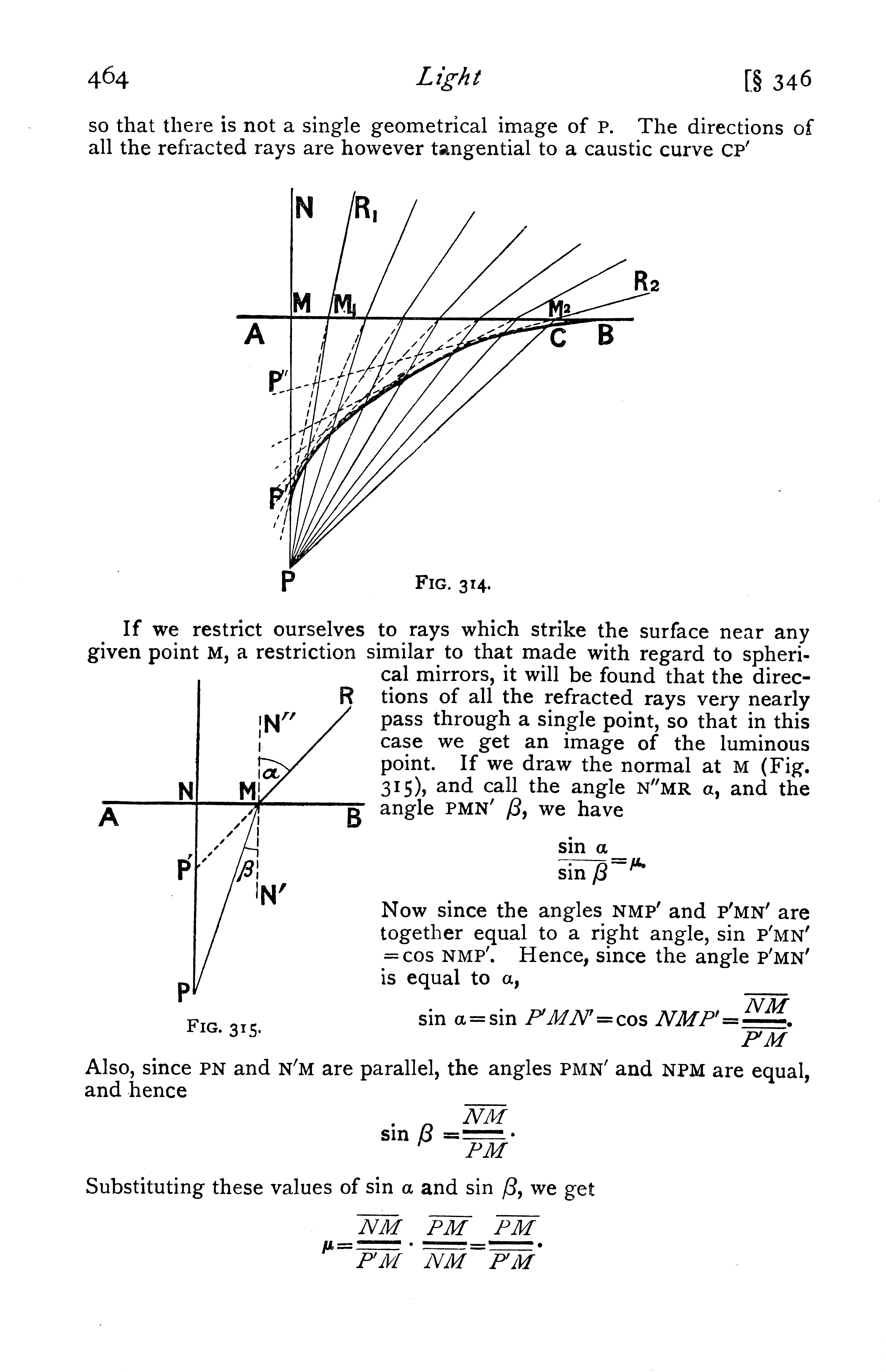}
\includegraphics[width=0.45\columnwidth]{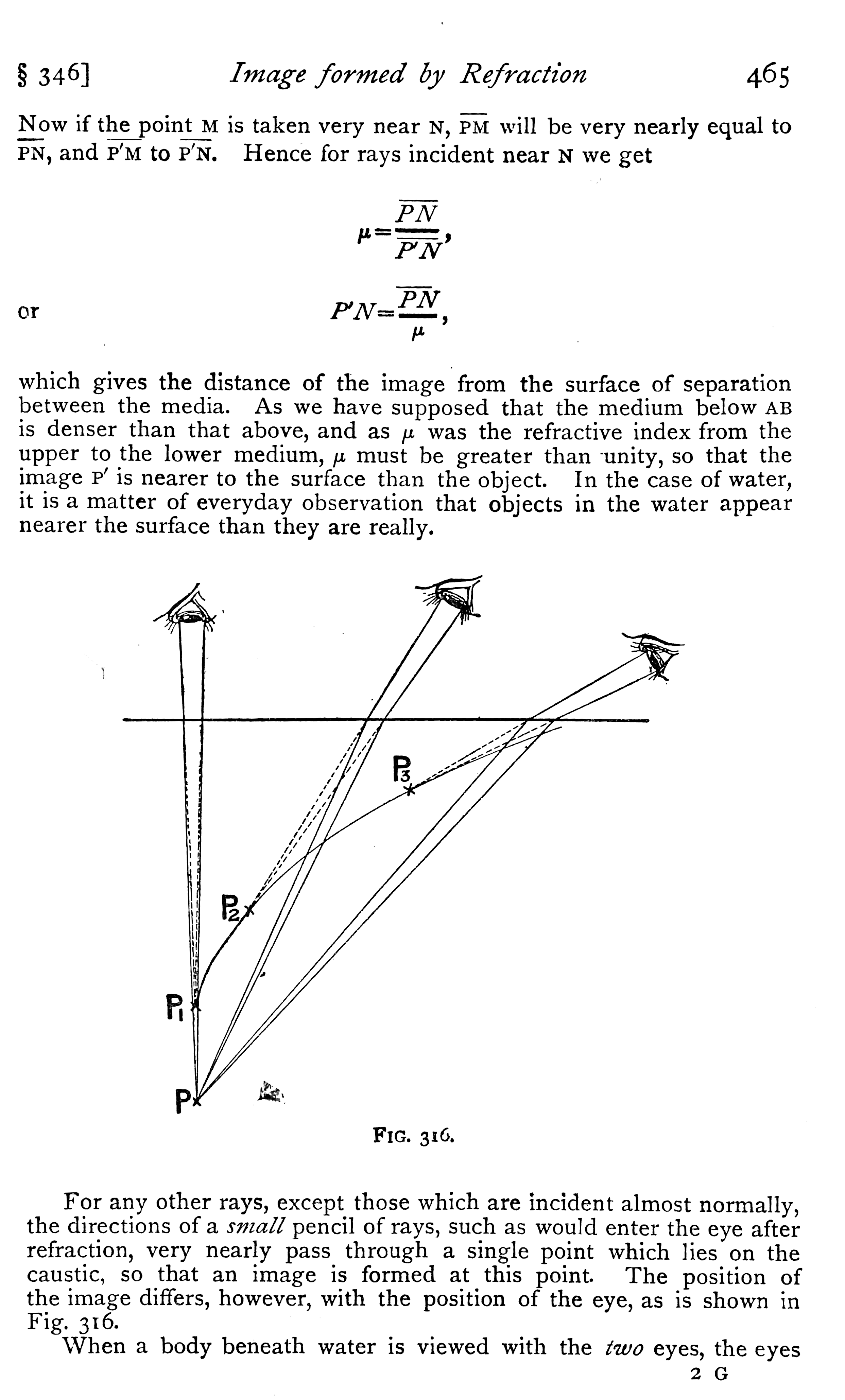}
\caption{Two pages from the book of
  Watson.\citep{watson1907} Figure 316, on the right, describes the
  trajectories of the rays coming from the source point $P$ which is
  immersed in the water. Depending on the height of the eye above the
  water, these rays seem to come from
  the points $P_1$, $P_2$, and  $P_3$. 
  This construction and its illustration, like many others, is purely
  2-dimensional. 
  However, as Feigenbaum discovered, to really capture the complete field of
  rays, the corresponding caustics, and the astigmatic effects of the
  two caustics, one needs to do a 3-dimensional calculation.
  There are then 2 points of large intensity ($H$ and $V$) as shown in \fref{fig:wcaustic}.}
\label{fig:watson}
\end{figure}

\begin{figure}[h!]
\centering
\includegraphics[width=\columnwidth]{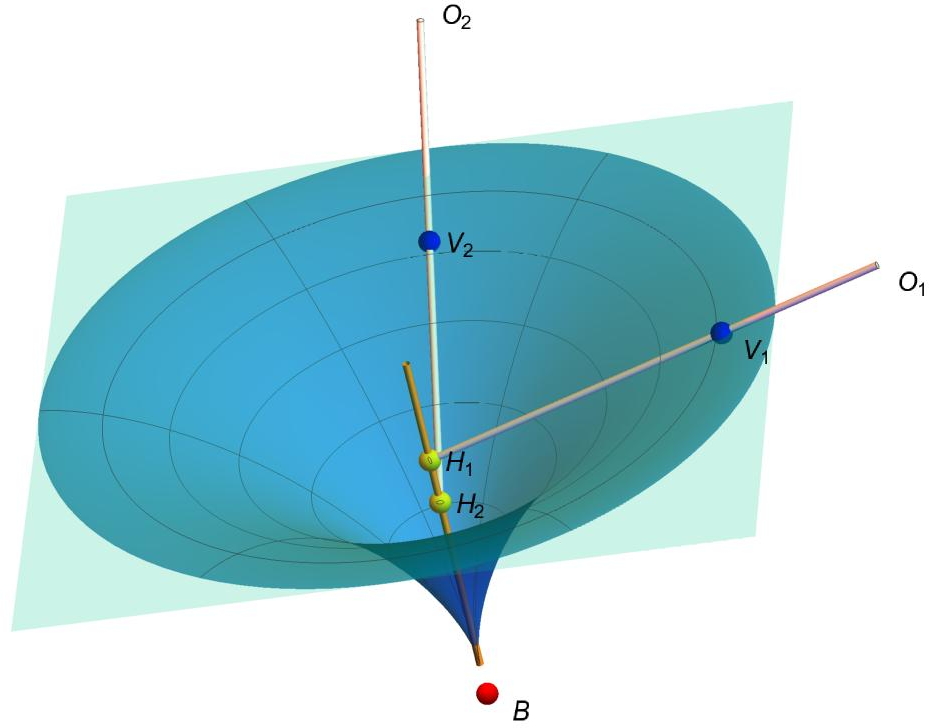}
\caption{(Color online) The transparent square surface
  represents the surface of the water. The red dot is the position of
  the light source. The blue funneled surface is the
  well-known caustic (now shown in 3D). (It does not extend to the red
  dot.) The dark yellow vertical line is the
  locus of the $H$ caustics. 
  The views of 2 observers, $O_1$ and $O_2$
  are shown. Each observer can see \emph{two} points of high
  intensity corresponding to the point $B$. The two points are called $V$
  respectively $H$.
  Note that $H_1$ and $H_2$ (yellow) slide on a vertical line, while the $V$'s
  (blue) lie on the  curved surface. This is where the observer will
  see the $V$ caustic, depending on height of the eye above the water.
  The $V$ were known to physicists, as in
  Watson.\cite{watson1907} The $H$ appears also in some references such as
  Quick.\cite{Quick2015}  The apparent motion of the ruler will appear
  because, depending on the angle at which a point appears in the
  water, its distance will vary (from on $V$ point to another).
  }
\label{fig:wcaustic}
\end{figure}

The setup  (\fref{fig:rulerinwater}) will be described in detail below, but
it is interesting to consider what was known in the XXth century about
looking into
water. An early reference is the 1907 book by
Watson,\citep{watson1907} of which pages are reproduced in
\fref{fig:watson}.
 One can clearly see that people realized at the time
that, depending on the height of the eye above the water, one sees the most intense
point at different depths in the water (see Fig.~316 of
Watson,\cite{watson1907} reproduced in \fref{fig:watson}). But 
authors before Feigenbaum seem to have overlooked that there are
\emph{two} caustic points, which are shown in \fref{fig:wcaustic}: The
well-known ones, denoted $V_1$ (respectively $V_2$), and the new ones
denoted $H_1$ (respectively $H_2$). This drawing shows
the point emitting light at $B$.

\section{What does one see?}
The fact that there are two candidates for an intense caustic point
raises now the important, and quite novel, question: Which of the two does one
see?

\begin{figure}[h!]
\centering
\includegraphics[width=2in]{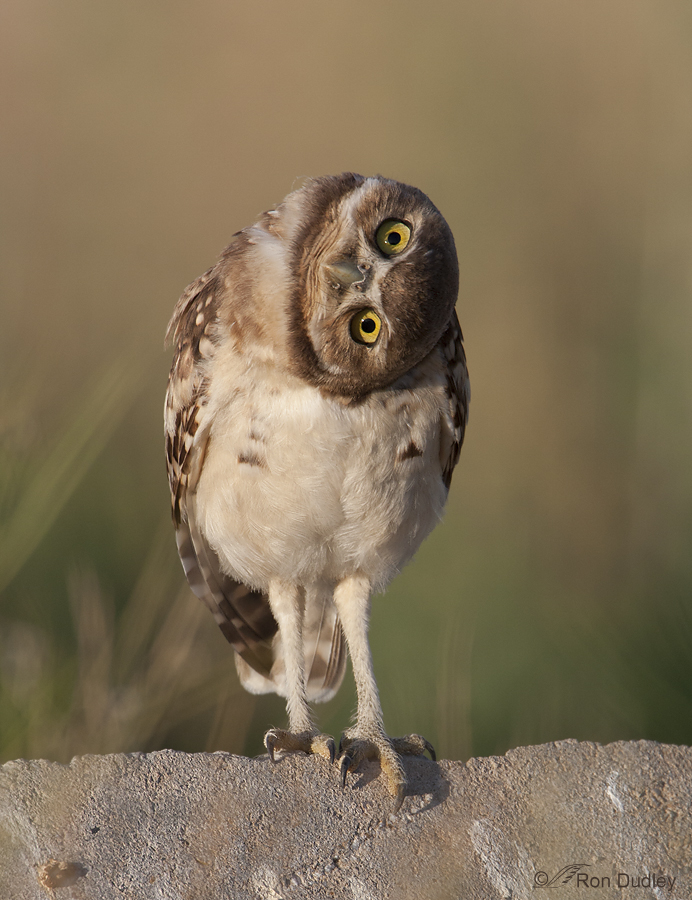}
\caption{(Color online) The meaning of tilting the head by 90
  degrees. Copyrighted photograph by Ron Dudley.
  }
\label{fig:cat}
\end{figure}

And here is the surprise: With the head upright, looking into the
water, we see the $H$ points (which seem to move when we move the head
up and down). This is what follows from the 2-dimensional theory as
shown in \fref{fig:watson}. However,
tilting one's head sideways (like the owl in \fref{fig:cat}) one prefers the $V$ points. The effect, if you do
the experiment, is that when you tilt your head, the bottom of the
ruler seems to move toward you. (I describe in more detail below what
you should expect, when I show how to best carry out the experiment.)

Experimentation in a class will show that about 80\% of
people see the effect. Note that the effect does \emph{not} depend on
binocular vision as you can easily check by closing one eye.

\section{Calculating the caustic}

\begin{figure}[h!]
\centering
\includegraphics[width=\columnwidth]{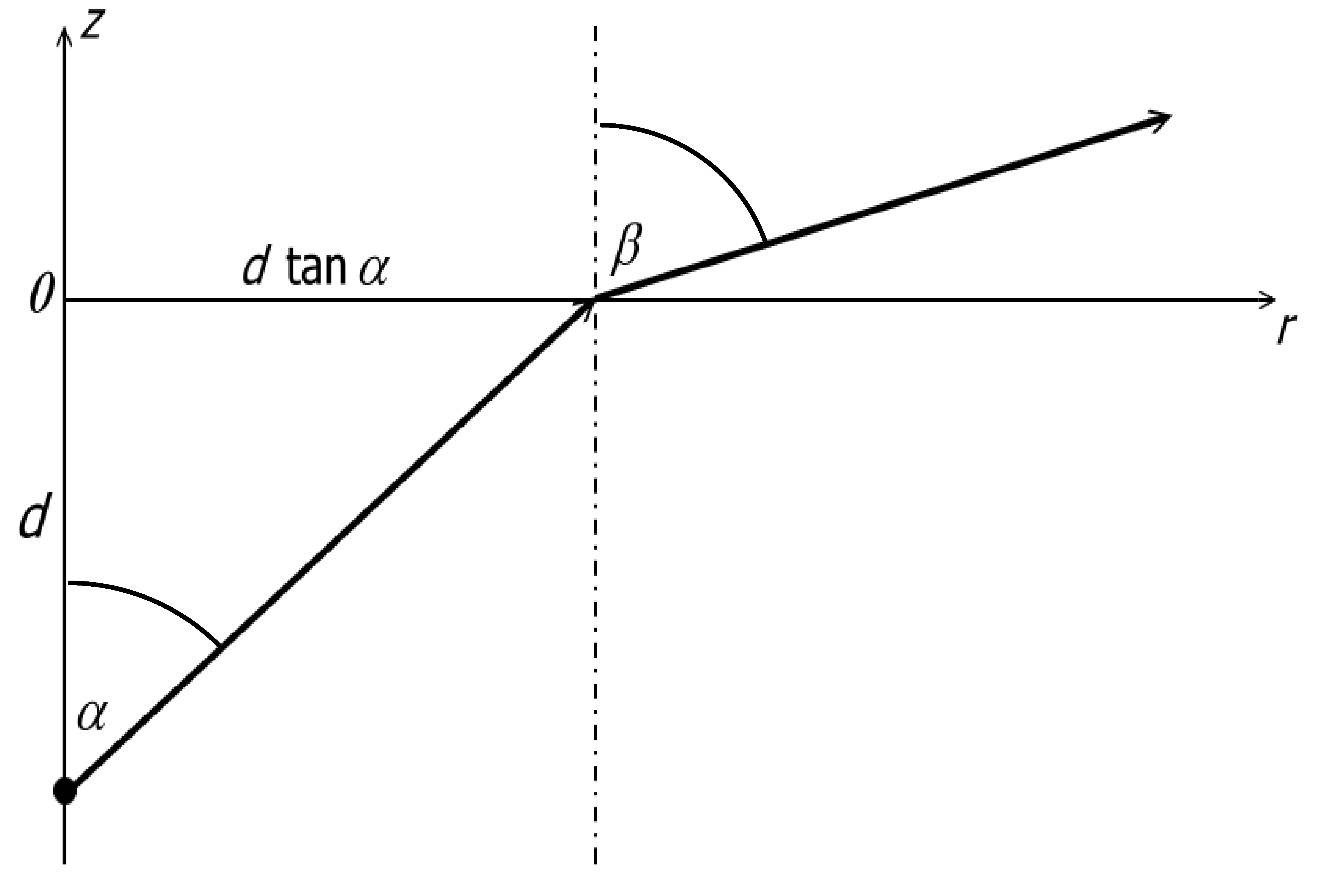}
\caption{Because the index $n=1.33$ of water is larger than that of
  air ($n=1$), an object at depth $d$ sends rays which exit the water
  at $z=0$ at the
  angle $\sin\beta=n \sin\alpha $. Note that this figure appears also
  as Figure 315 in \fref{fig:watson}.
  }
\label{fig:snell}
\end{figure}

While much of this material is already explained in\cite{watson1907} and in
many textbooks, we need to repeat it here, so that the reader
understands how the second caustic, the $H$ caustic, appears. And to
understand its astigmatism, one really needs to do the 3d calculation. While
it is very natural to do the 2d calculation as in Watson,\cite{watson1907}
it is just not enough, because the astigmatism at $H$ extends in the
direction orthogonal to the plane $(r,z)$ of \fref{fig:snell}.

The calculation starts with the index of water, $n\sim 1.33$. By
Snell's law, \fref{fig:snell}, $\sin\beta=n \sin\alpha $.
%It is important, and different from the classical literature, that one
%performs the calculation in 3 dimensions, not two (because the
%astigmatism of the $H$ image only appears in the 3rd dimension).
We use complex coordinates in the plane $z=0$ that defines the surface
of the water, and write $(x,y)=x+i
y=r e^{i\phi}$.
Any point $\x\in{\mathbb R}^3$ can then be conveniently be written as
$\x=(re^{i\phi},z)$.
With this notation, a point on the outgoing ray can be represented as
$\p=\p_0+\ell\, \hat \t\in {\mathbb R}^3$, with $\ell\in{\mathbb R}$ and 
\begin{equa}
  \p_0&=(D\frac{t}{\mu}e^{i\phi},0)~,\qquad
  \hat \t=\frac{(te^{i\phi}/\mu, \sqrt {1-t^2})}{\sqrt{1+t^2/\nu^2}}~.
\end{equa}
(The point $\p_0$ is in the plane $z=0$ and $\hat \t$ is the unit
vector along the outgoing ray.)
Here, $\nu=\sqrt{n^2-1}$, $\mu=\nu/n$, $t=\nu \tan \alpha $, and
$D=d/n$. The angle $\phi$ is the angle in the $(x,y)$ plane
(perpendicular to the $z$-axis).
The caustic is that surface which is tangent to $\hat\t$ (at every
point).
Eliminating
$\ell$ and expressing the result $\underline x=(x,y)$ in terms of $z$  one gets
\begin{equ}\label{eq:ray}
\underline x= \underline x(z)=
\frac{t}{\mu}e^{i\phi}\left(D+\frac{z}{\sqrt{1-t^2}}\right)\in
{\mathbb R}^2.
\end{equ}
Note that $\underline x$ is parameterized by $t$ and $\phi$, with $D$
and $\mu$ being fixed quantities related to the depth of the source point
and the index of refraction of water.

The caustics are now found by requiring that the differential of
\eref{eq:ray} vanishes, because we want the caustics to be tangent to
the rays. This means we have to calculate the
derivatives with respect to $t$ and $\phi$. These two variables are
angles: $t=\nu \tan \alpha $ is related to the vertical angle $\alpha $, while
$\phi$ is the angle in the $(x,y)$ plane.
The differential is equal to
\begin{equa}[equ:222]
  &\frac{it e^{i\phi}}{\mu}\left(D+\frac{z}{\sqrt{1-t^2}}\right)\cdot
  \d \phi\\+&\frac{e^{i\phi}}{\mu}\left(D+\frac{z}{\sqrt{1-t^2}}
  +\frac{t^2 z}{(1-t^2)^{3/2}}\right) \cdot \d t~.
\end{equa}
We want this to vanish. 
The
well-known classical caustic surface is the one for which $\d \phi=0$
(which means that the coefficient of $\d t$ must vanish). This
produces the point
\begin{equ}
  r_V=\frac{D}{\mu}t^3 ~\text{  and  } z_V=-D (1-t^2)^{3/2}~.
\end{equ}
(First solve for $z_V$ and substitute into the coefficient of $\d\phi$
to obtain $r_V$.)
These  are the points indicated by $V$ in \fref{fig:wcaustic}. They
are unsharp in the horizontal direction because $\d\phi$ is a
variation in the horizontal plane $(x,y)$.
Similarly, setting $\d t=0$ will produce the $H$ points
\begin{equ}
  r_H=0 \text{  and } z_H=-D\sqrt{1-t^2}~,
\end{equ}
which are vertically unsharp, as $\d t$ is a variation of the vertical
angle $\alpha $.
The cusp (of the $V$ caustic) in \fref{fig:watson} obeys the equation
\begin{equ}
  (\mu r)^{2/3} + (-z)^{2/3}=D^{2/3}~.
\end{equ}
Note that the two caustic points $V$ and $H$ coincide when $t=0$, \ie
when one looks
vertically down into the water to the source point. This point has
perfect focus, since the two caustics coincide.

\begin{figure}[h!]
  \includegraphics[width=\columnwidth]{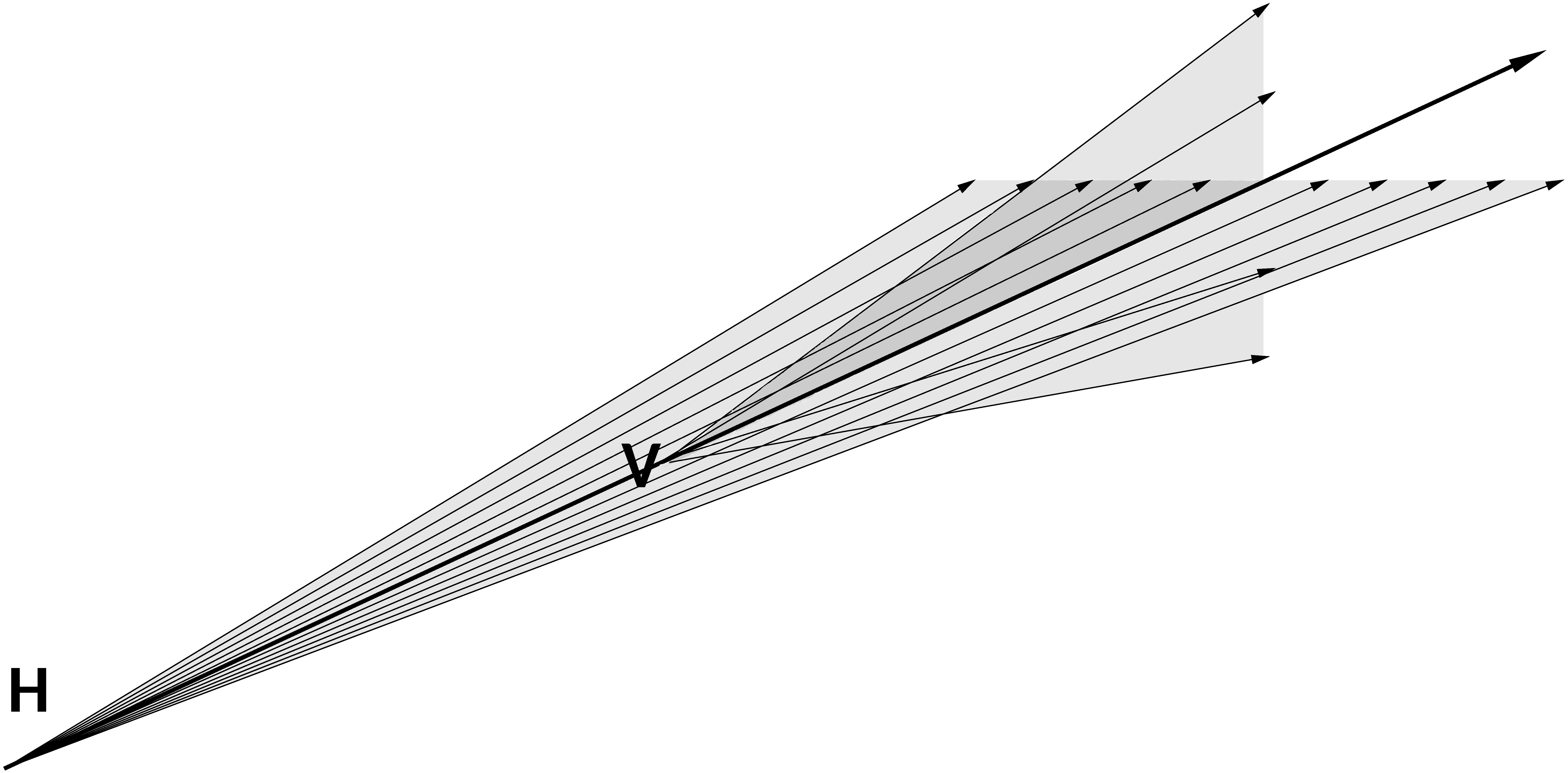}  
  \caption{Illustration, modified from Feigenbaum, of  two orthogonal fans of
    rays leaving points $H$ and $V$.}\label{fig:fan}
\end{figure}

Our description of $\d \phi=0$ and $\d t=0$, 
shows that the rays coming out of $H$ and $V$ form two orthogonal fans,
as illustrated in \fref{fig:fan}:
The fan of rays coming out of point $H$ is \emph{horizontal}
while the fan of rays coming out of the $V$ points is \emph{vertical}
(relative to the surface of the water). 

I insist again: The novelty of this approach is to have performed  the
calculation in 3 dimensions, not just in the $(r,z)$ plane. Without
this extension, the combination of the $H$ and the $V$ caustics will
not be discovered in just one differential, namely \eref{equ:222}. This is
Feigenbaum's mathematical contribution to the question of imaging.

I want to end this section with some historical remarks.
The existence
of the $H$ points (in addition to the $V$ points) appears in Kinsler,
Bartlett, Horvath.\cite{Kinsler1945,Bartlett1984,Horvath2003} An interesting sequence of
papers is Nassar's view on ``apparent depth,''\cite{Nassar1994} on which Bartlett\cite{Bartlett1995} and Mosca\cite{Mosca1995} commented. In particular, Bartlett
cites Sears,\cite{Sears1949} which contains a calculation of the
astigmatics (on page 42).

A more recent, and very complete reference of importance in the
subject is \cite{Quick2015} which discusses both caustics
(unfortunately in German).
These authors  observed that both $H$ and $V$
are on the same line of sight as illustarted for the two observers in
\fref{fig:wcaustic}.  

An important reference with more mathematical inclination is Berry,\cite{berry1981} which shows clearly, in
Fig.~28 on page 484 a
sketch of what Feigenbaum showed in \fref{fig:wcaustic}. But there is
no mention of the role of astigmatism.

So, the
novelty of Feigenbaum's approach is to connect the existence of the
two points with the difference in astigmatic direction. To discover
this, the 3d calculation is essential. And the change of astigmatic
direction from vertical to horizontal (at $H$, resp.~$V$) is
the cause that the ruler seems to move when the head is tilted.

\section{The role of the eye}

So far, I have mainly concentrated on the optics of looking into
water. But to really understand what one sees, 
I need to explain certain properties
of the human eye.

Our eyes are not perfect. Some people
are near- or far-sighted\cite{wiki:nearsighted,wiki:farsighted} or
cannot accommodate\cite{wiki:accomodation} without glasses. These imperfections appear when the focus of
what we see lies in front of or behind the retina (which is the capturing
device in our eyes). Since the lens and the retina are not a perfect
camera, the brain will correct some of the errors, if they are not too
large.

There are other imperfections of the eye, and the one of interest to
us is astigmatism.\cite{wiki:astig} This notion is used when the image of a point is
mapped to a small line segment on the retina. Depending on how large
this line segment is, glasses must be made to correct for this, so that
the retina gets to perceive a perfect point.

Ophthalmologists know that the
necessity for optical correction depends, astonishingly, on the \emph{direction}
of the small line segment. In fact, if the line segment is vertical
(called ``with the rule'' (WTR)), a correction is much less needed than
if it is horizontal, (called ``against the rule'' (ATR)).\cite{wiki:astig} For
  the frequency of astigmatic prevalence, see Nemeth et al.\cite{wtr} The
evolutionary origin of this asymmetry seems not known: Is it gravity,
looking at faraway things at the horizon, or a remnant that mammals
descend from aquatic animals? Still, for our experiment, this
asymmetry of dealing with unsharp images is crucial:
After all, as I have shown above, both images, the $H$ and the $V$,
are unsharp, since they move perpendicular to the fans of
\fref{fig:fan}, when the eye is moved perpendicularly to the
fans. But, since the eye has a non-vanishing opening, this moving
effect happens even if the eye is in a fixed position. (It is like
moving the eye by its opening, about 5mm.)
The $H$ caustic produces an
image which is vertically unsharp (relative to the surface of the
water) while the $V$ image of any point is spread horizontally, when
it reaches the eye.

Since, as I explained above, our eyes correct more easily vertical
unsharpness (WTR), we will preferentially focus on the $H$ caustic
when the head is upright. Tilting the head by
90 degrees has the effect of switching ``vertical'' and ``horizontal.''
And now the eye-brain system will prefer to focus on $V$. And, as I
said, we do not understand the reason for the preference of WTR over
ATR. The experiment of the ruler in the water can thus be understood in
terms of this preferred focus. It is not an effect of binocular
vision, as you should check by closing one eye.

\section{The experiment}

\begin{figure}[h!]
\centering
\includegraphics[width=2in]{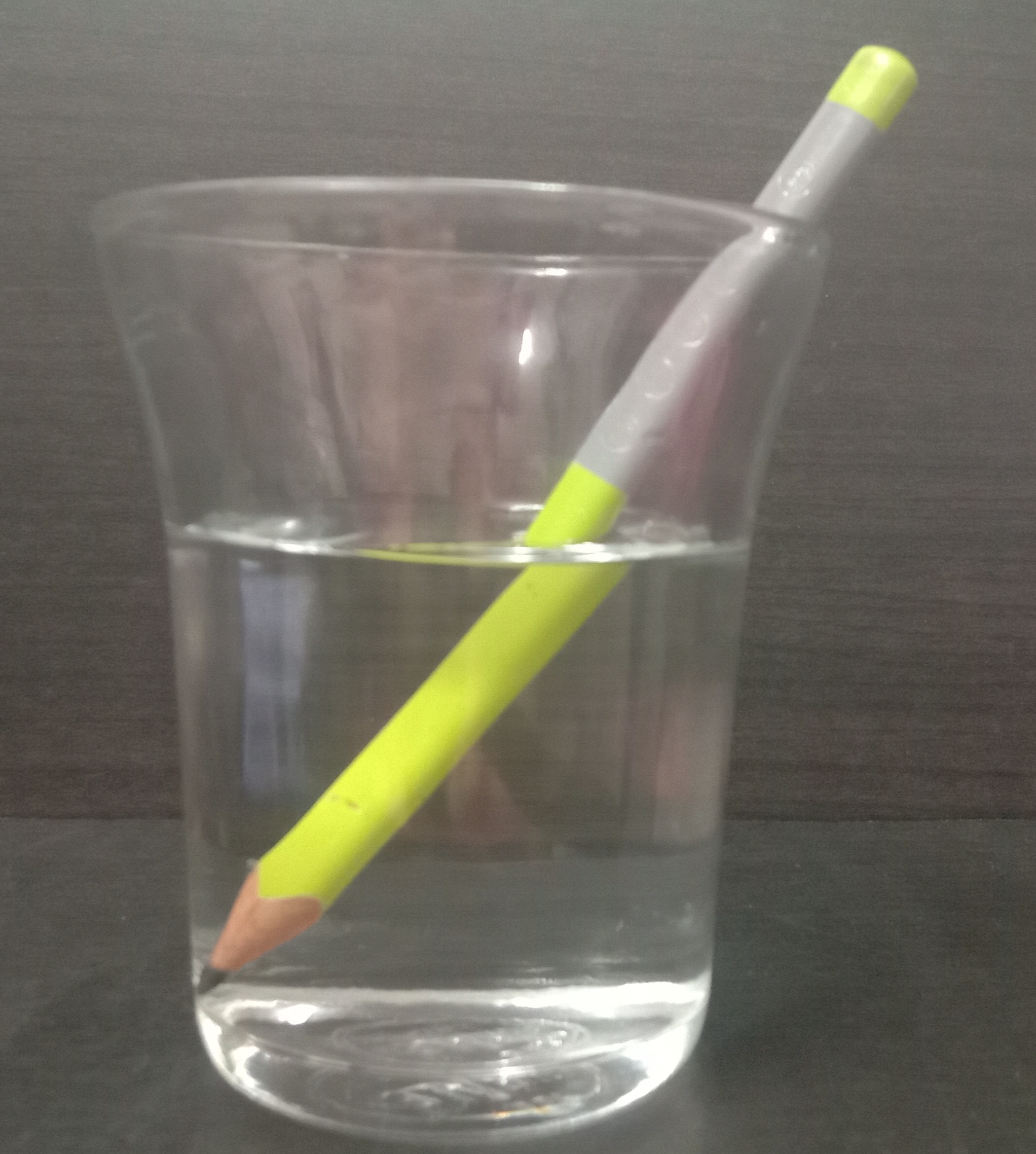}
\caption{(Color online) The ``classical'' broken pencil. Picture by
  Kunal B Mehta (Wikimedia commons, cropped from original).
  }
\label{fig:pencil}
\end{figure}

One should note that this experiment is not
the well-known phenomenon of the ``broken pencil''
of \fref{fig:pencil}, which appears when one looks \emph{through} the water (and
air) and not \emph{into} the water as I will describe.

\begin{figure}[h!]
\centering
\includegraphics[width=3in]{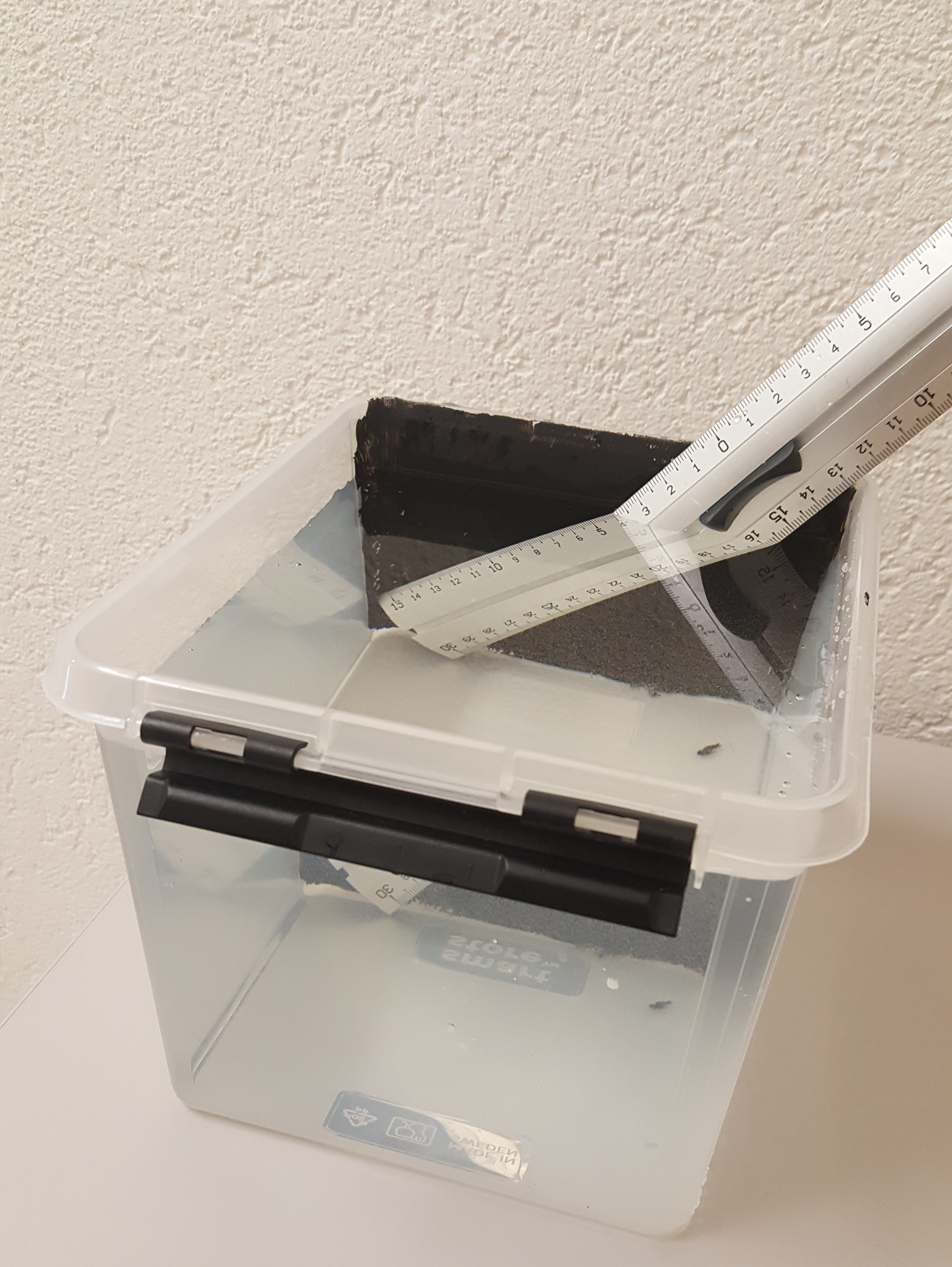}
\caption{(Color online) How to put the ruler into the water. The scale on the ruler
  is in cm. To look at the ruler, lower your eye to the upper rim
  (above the black handle), as close to the surface of the water as possible.
  }
\label{fig:rulerinwater}
\end{figure}

A good setup is a plastic container, of dimension about
$20\times15\times15$ cm,
placing the ruler at the far end, parallel to the (blackened)
wall, as in \fref{fig:rulerinwater}. The container should be filled as
much to the rim as possible.

\begin{figure}[h!]
\centering
\includegraphics[width=\columnwidth]{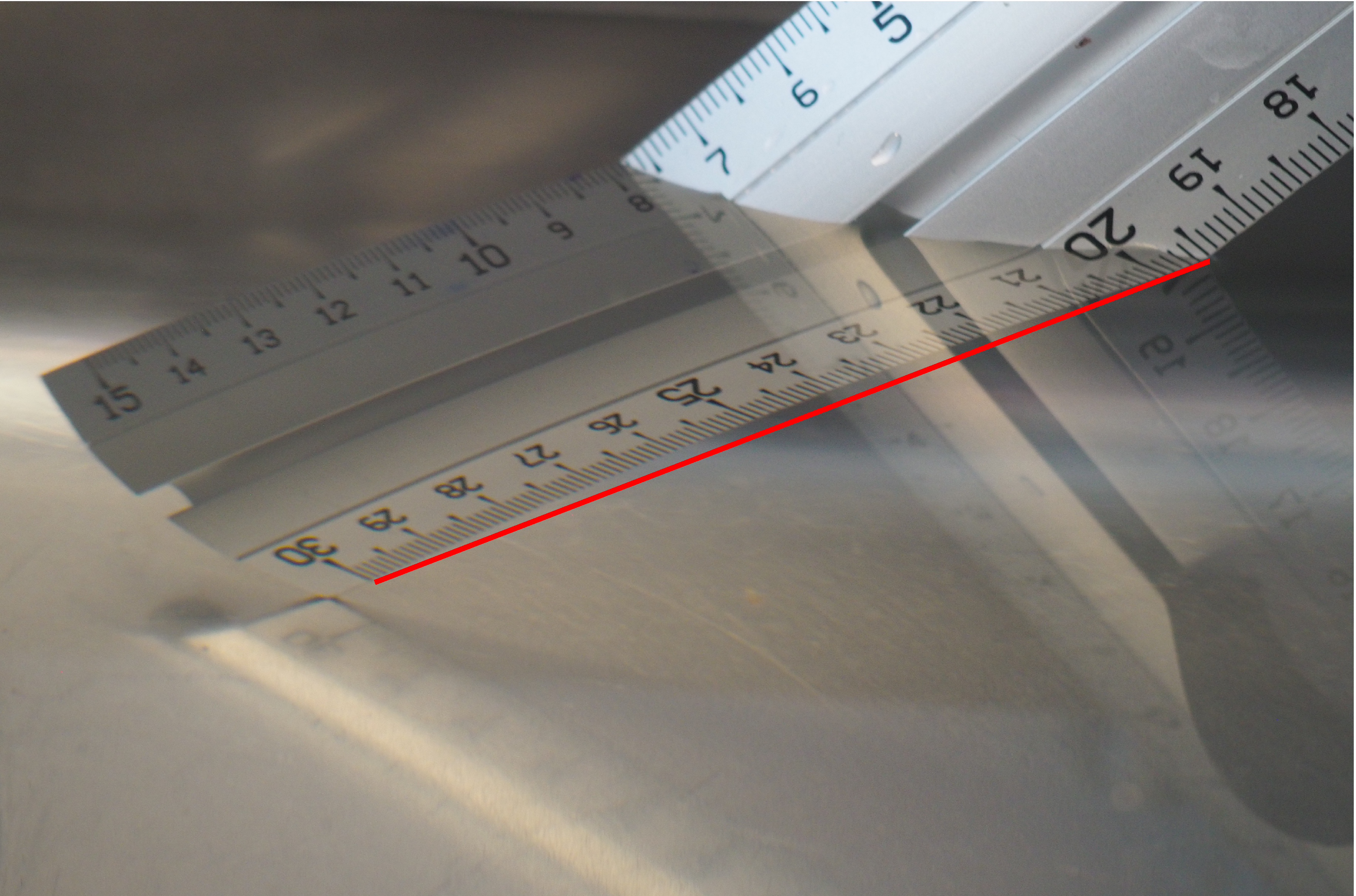}
\caption{(Color online) A photograph of the ruler, looking into the water from above
  (at a flat angle). The ruler looks clearly curved.
  }
\label{fig:pencil3}
\end{figure}

The viewer should look into the water through the surface at the
flattest possible angle for which one still sees the bottom of the ruler.
You can then see that the ruler is not straight, but slightly curved, as
in \fref{fig:pencil3}. 
This is already described in Watson,\citep{watson1907}
as shown in
\fref{fig:watson}. Since the points of vision $P_1$, $P_2$ ,$P_3$
``slide down'' as a 
function of the angle at which  you look into the water, inclined
straight objects appear curved. This was certainly known to
physicists at the end of the 19th century.

However, the new effect, discovered by Feigenbaum, appears
if you  tilt your head (keeping the eye in the same position
relative to the box). What I mean by tilting is 
shown in
\fref{fig:cat}.

Then the bottom corner of the ruler (the point near the
'30' graduation of the ruler) seems to move towards you, quite a bit. The eye re-focuses as if on an object coming
closer. And the top of the ruler,
at '15' graduation across from the '30', seems to move less toward you (since it
is less deep in the water). This is the $V$ caustic you are seeing.

As Feigenbaum noted, this is what happens  ``naturally.''  However, if you
force yourself and change focus willfully, you can choose to see the
other caustic.

A final remark: Note that no photograph can capture this effect,
because the eye has a second property: It focuses on what the brain
can sharpen best. And in this case, the upright eye will focus on the
$H$ position. This is because the image is vertically unsharp, and the
eye-brain system can more easily make a sharper image in contrast to
horizontal blurriness, see \fref{fig:wcaustic}. The
tilted eye will focus on the $V$ caustic, which slides along the
surface. An amusing consequence of the ambiguity on the choice of
focus appears when you want to photograph the scene with a modern camera: Often, the
auto-focus will have difficulty ``deciding'' what to focus on.

\label{sec:sascha}
I will end by adding a suggestion of a variant setup by one of the referees.
\begin{quote}
``I drew a pattern of horizontal and
 vertical lines on a vertical plane and submerged that vertical plane
 under water. With my eye close to the waterline, I managed to get a
 vertical line in focus ($H$ image) for the untilted head, and a
 horizontal line in focus ($V$ image) for the tilted head. However, I
 also managed to make the opposite observation: I was able to get a
 horizontal line in focus ($V$ image) for the untilted head, and a
 vertical line in focus ($H$ image) for the tilted head. Finally, I also
 tried a motion from untilted to tilted head, and I was able to keep
 one type of line (horizontal or vertical) in focus, depending on my
 will. Therefore, I conclude that both types of image are successively
 visible with monocular vision, independent of head orientation. It
 solely depends on the intention of the viewer.''
\end{quote}
 The interested
 person can test this ambiguity. What this says is that a willful 
 change of focus, allows one to
 see the picture which is unsharp in the
 ``wrong'' direction.  As Feigenbaum points out, slight horizontal motion of the
 head makes the $H$ image disappear. Like in the case of the cylinder
 of \fref{fig:anacyl} the image of the square grid appears on a curved
 surface--the ``viewable surface''--which is again difficult to compute.

\section{Conclusion}

What should one carry home from this? To me, the study of Feigenbaum
teaches us that it is worthwhile doing a careful calculation,
well-adapted to the problem. But it also tells us that it is good to
think beyond the problem at hand. While we still do not understand why
Nature has given preference to automatically correcting vertical
astigmatism, we certainly can see how suddenly the study of a simple
physical effect can inspire astonishing connections between two
seemingly unrelated disciplines, in this case optics and ophthalmology.

\section*{Acknowledgment}
I am grateful that the late Mitchell Feigenbaum discussed with me his
book for about 15 years, and that he regularly shared with his
insights and the  many
versions of the manuscript.
I have profited from many discussions with Gemunu Gunaratne.
I thank the referees for many helpful remarks and bibliographical
help. In particular, Sascha Grusche suggested a variant of the
experiment which I incorporated in the text. Financial
support is from ERC advanced grant Bridges. This publication was
produced within the scope of the NCCR SwissMAP which is funded by the
Swiss National Science Foundation.

\end{document}